\documentclass{article}

    \usepackage[preprint]{neurips_2024}

\usepackage[utf8]{inputenc}
\usepackage[T1]{fontenc}
\usepackage{hyperref}
\usepackage{url}
\usepackage{booktabs}
\usepackage{amsfonts}
\usepackage{nicefrac}
\usepackage{microtype}
\usepackage{xcolor}

\usepackage{hyperref}
\hypersetup{
    colorlinks=true,
    linkcolor=blue,
    filecolor=blue,
    urlcolor=blue,
    citecolor=blue,
    }
\usepackage{graphicx}
\usepackage{array}
\usepackage{makecell}
\usepackage{longtable}
\usepackage{listings}
\usepackage{adjustbox}
\usepackage{svg}
\usepackage{listings}
\usepackage{todonotes}

\title{Generative UI: LLMs are Effective UI Generators}

\author{
  \textbf{Yaniv Leviathan} \quad \textbf{Dani Valevski} \quad \textbf{Matan Kalman} \quad \textbf{Danny Lumen} \\[4pt]
  \textbf{Eyal Segalis} \quad \textbf{Eyal Molad} \quad
  \textbf{Shlomi Pasternak} \quad \textbf{Vishnu Natchu} \quad \textbf{Valerie Nygaard}  \\[4pt]
  \textbf{Srinivasan (Cheenu) Venkatachary} \quad
  \textbf{James Manyika} \quad
  \textbf{Yossi Matias} \\[6pt]
  Google Research\\[4pt]
  \scriptsize{\texttt{\{leviathan, daniv, matank, dwasserman, eyalis, moladeyal,}} \\ \scriptsize{\texttt{spasternak, vnatchu, vnygaard, vsri, jmanyika, yossi\}\texttt{@google.com}}}
}

\begin{document}

\maketitle

\begin{abstract}
AI models excel at creating content, but typically render it with static, predefined interfaces.
Specifically, the output of LLMs is often a markdown ``wall of text''.
\emph{Generative UI} is a long standing promise, where the model generates not just the content, but the interface itself.
Until now, Generative UI was not possible in a robust fashion.
We demonstrate that when properly prompted and equipped with the right set of tools, a modern LLM can robustly produce high quality custom UIs for virtually any prompt.
When ignoring generation speed, results generated by our implementation are overwhelmingly preferred by humans over the standard LLM markdown output.
In fact, while the results generated by our implementation are worse than those crafted by human experts, they are at least comparable in 50\% of cases.
We show that this ability for robust Generative UI is emergent, with substantial improvements from previous models.
We also create and release \emph{PAGEN}, a novel dataset of expert-crafted results to aid in evaluating Generative UI implementations, as well as the results of our system for future comparisons.
Interactive examples can be seen at \href{https://generativeui.github.io}{generativeui.github.io}.
\end{abstract}

\begin{figure}[ht!]
  \centering
  \begin{adjustbox}{center}
    \includegraphics[width=1\textwidth]{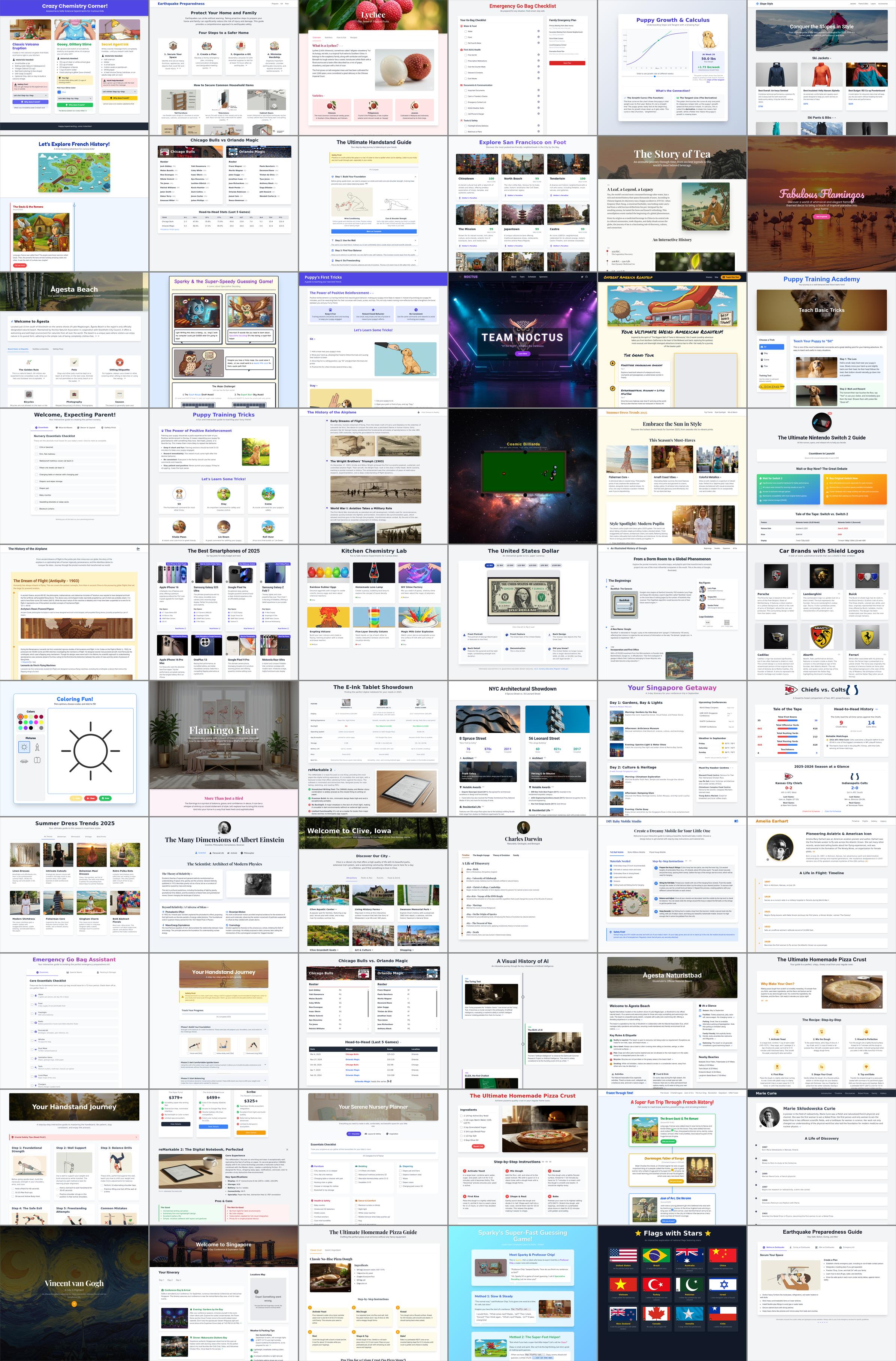}
  \end{adjustbox}
  \caption{Results from our implementation (see \href{https://generativeui.github.io}{generativeui.github.io}).}
\end{figure}

\section{Introduction}
\label{section:intro}

AI models today generate \emph{content}: text, code, images, videos, etc. However, the results of these powerful tools are often presented using hard-coded and pre-designed user interfaces. \emph{Generative UI} is a new modality where the AI model generates not only content, but the entire user experience. This results in custom interactive experiences, including rich formatting, images, maps, audio and even simulations and games, in response to any prompt (instead of the widely adopted ``walls-of-text'').

\paragraph{An instant AI team for each prompt.} Today, rich visual interfaces exist for common user journeys.
Specifically, teams composed of product managers, UX designers, and engineers work for extended periods of time to build amazing rich experiences for \emph{broad prompt categories}, shared by many users.
Generative UI enables us to spin up an instant (AI-based) product management, UX design, and engineering teams, to build an interactive experience, over the course of a minute, for a specific prompt.
While not as competent as human experts, Generative UI enables custom experiences for \emph{any prompt}.

At present, the prevalent UI for interacting with LLMs is a markdown-based chat interface.
Specifically, the model outputs markdown (that can include heading, emojis, tables, code-blocks, etc.).
These are significantly easier for humans to consume than raw text, yet results created by our Generative UI implementation are overwhelmingly preferred over both (see Table~\ref{tab:sxs-comparison}).
To evaluate our implementation of Generative UI we use human rater preference compared to a set of baselines. We collect and make available \emph{PAGEN} (see Section \ref{section:pagen}), a dataset of pages made by human experts for custom prompts.
While the expert-made pages are broadly preferred to those from our system, we show that for the first time we get comparable results on a large fraction of prompts.
See Section \ref{section:results} for details, Appendix \ref{appendix:examples} for screenshots, and \href{https://generativeui.github.io}{generativeui.github.io} for interactive results from our system.

\section{Method}
\label{section:method}

\begin{figure}[h]
  \centering
  \begin{adjustbox}{center}
    \includegraphics[width=\textwidth]{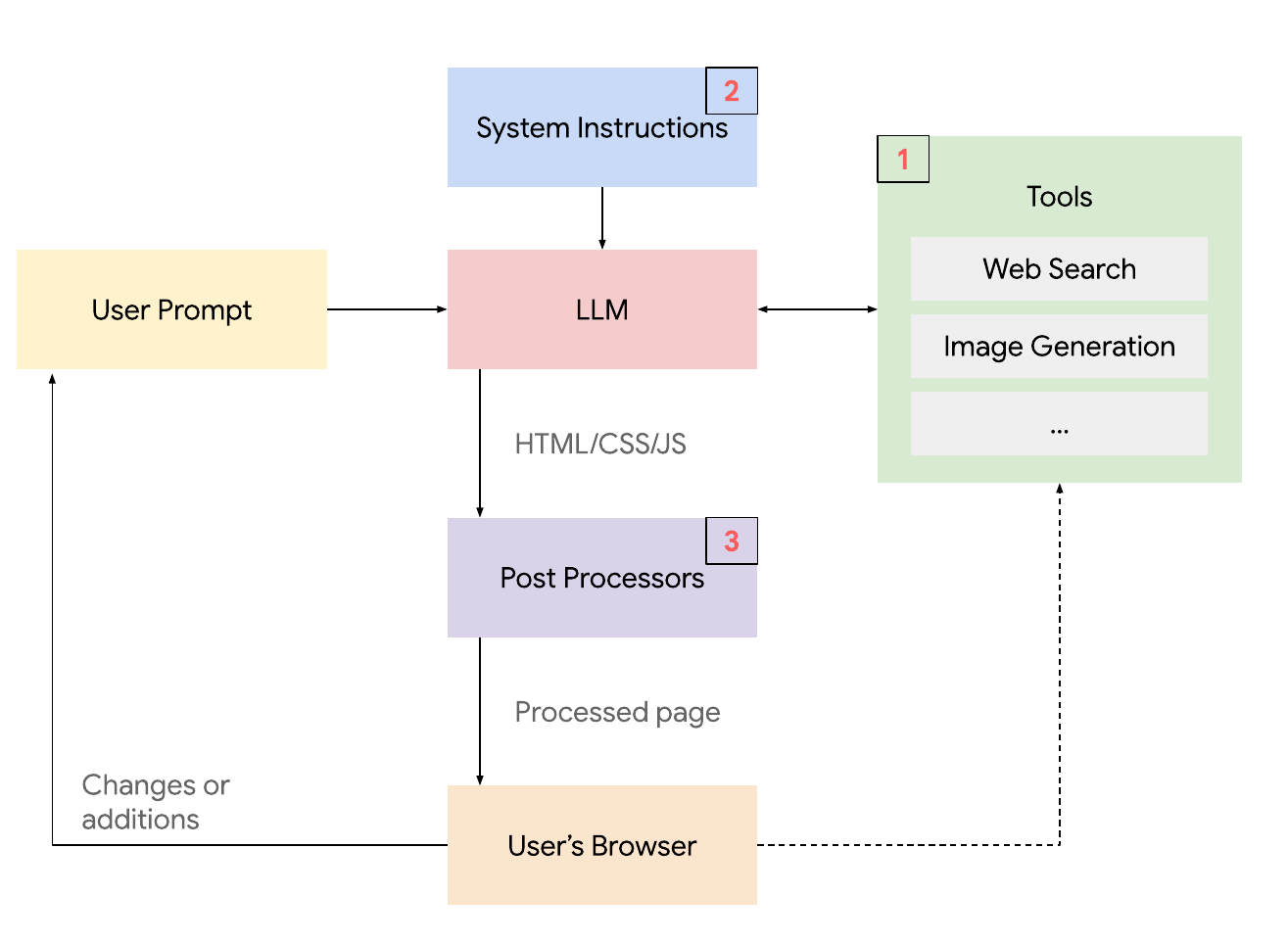}
  \end{adjustbox}
  \caption{A high level system overview.}
  \label{fig:fig1}
\end{figure}

Our Generative UI implementation outputs a single fully-generated web page and a set of accompanying assets, such as images.
The page is rendered as-is on the user's browser.
See Figure \ref{fig:fig1} for a high level overview of the system.

As depicted in Figure \ref{fig:fig1}, we employ 3 main components:
\begin{enumerate}
    \item A server exposes several endpoints enabling access to key tools, such as image generation and search.
The results can be made accessible to the model (increasing quality) or sent directly to the user's browser (increasing efficiency).
    \item Carefully crafted system instructions. These in turn include: (1) the goal (2) planning and thinking guidelines, (3) examples, and (4) a large set of technical instructions including formatting guidelines, tool endpoints manual, and tips for avoiding common errors. These contribute to the quality of the generated results (see Appendix \ref{appendix:the prompt} for an illustrative prompt from an early research prototype).
    \item A set of post-processors. These lightweight components address a set of remaining common issues. Additional post processors deal with error reporting and page analysis. See Appendix~\ref{appendix:post-processor}.
\end{enumerate}

\subsection{Consistent Styling}
\label{section:consistent styling}

\begin{figure}[h]
  \centering
  \begin{adjustbox}{center}
    \includegraphics[width=1\textwidth]{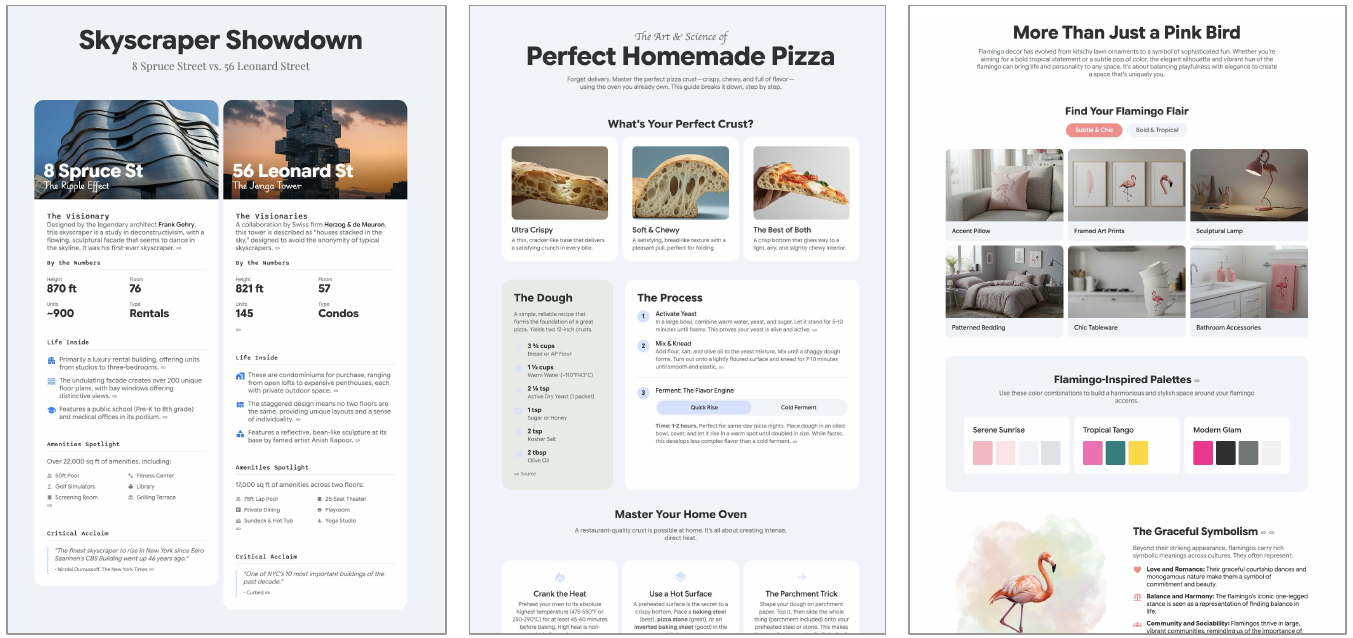}
  \end{adjustbox}
  \caption{Screenshots of Generative UI results with ``Classic'' styling.}
  \label{fig:classic_styling}
\end{figure}

\begin{figure}[h]
  \centering
  \begin{adjustbox}{center}
    \includegraphics[width=1\textwidth]{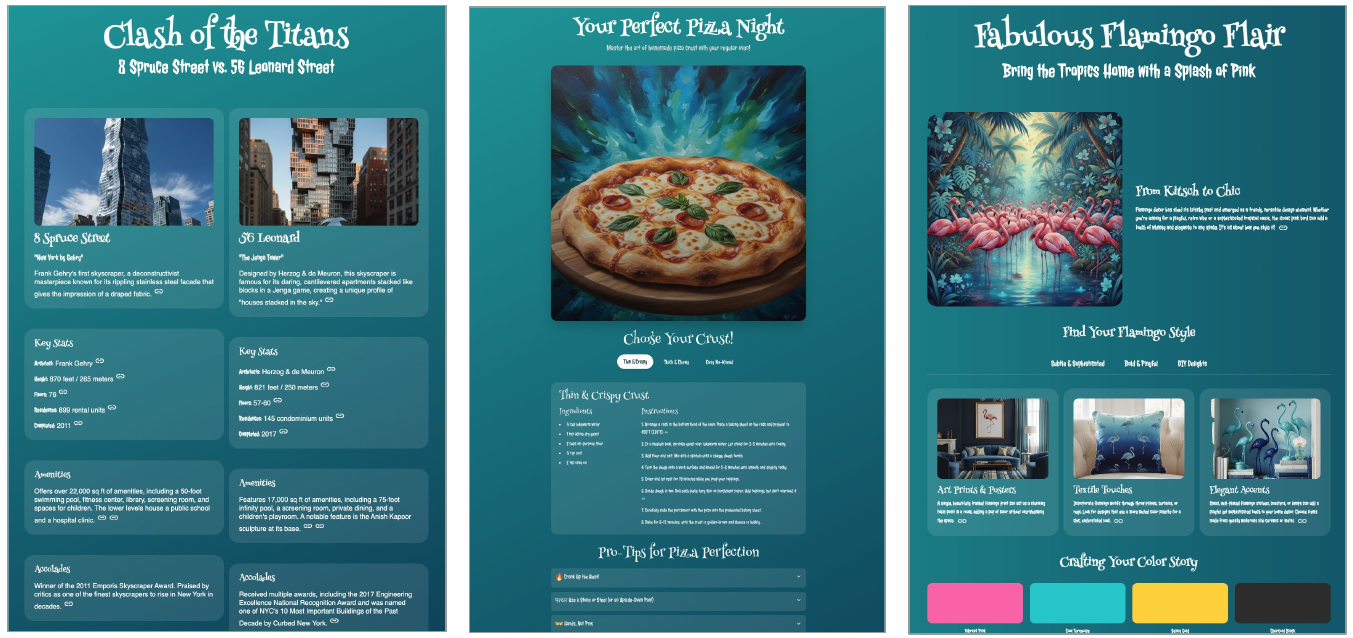}
  \end{adjustbox}
  \caption{Screenshots of Generative UI results with ``Wizard Green'' styling.}
  \label{fig:retro_styling}
\end{figure}

If desired, our setup allows producing results using a specific style and increased visual consistency across generations. This is done via small changes to the system instructions. Specifically, we experimented with replacing the short ``Style'' section in our prompt with more detailed variants (which we call ``Classic'' and ``Wizard Green''), specifying colors, fonts, etc. We observe that indeed the generated results follow these styles. Interestingly, the model automatically adapts all elements, including e.g. the generated images and icons to the desired styles. See Figures \ref{fig:classic_styling} and \ref{fig:retro_styling}.

\section{Results}
\label{section:results}

We evaluate user preference across several different result formats: a custom website crafted for the prompt by a human expert (see Section \ref{section:pagen}), the top Google Search result for the query, text (LLM output without markdown), standard LLM output (in markdown format),
and our Generative UI implementation.
We randomly sampled 100 prompts from LMArena \citep{chiang2024chatbotarenaopenplatform} (and excluded 8 queries, see Appendix~\ref{appendix:data collection}) and collected pairwise preferences from human raters, sending each result to 2 raters.
Generation time is not a factor in the evaluation and we show raters pre-cached results.
We ask the human raters to rate on a 3 point scale: Left Preferred, Neutral, Right Preferred.
In addition to the LMArena prompts, we also created a custom prompt set composed of information seeking prompts (see Appendix~\ref{appendix:info-seeking-prompts}). We evaluated on both sets using the same methodology.
See Appendix \ref{appendix:examples} for selected example generations from our Generative UI implementation.

Tables~\ref{tab:elo-scores}~and~\ref{tab:sxs-comparison} show the resulting ELO scores and side-by-side user preference for each of the UI modalities for the prompts from LMArena.
Generative UI obtains an ELO score of 1736.2, indicating a strong user preference over all other formats, except human experts.
Notably, when compared to Markdown UI - the next best method, Generative UI is preferred 82.8\% of the time.
See Appendix Tables \ref{tab:elo-scores-pagen} and \ref{tab:sxs-comparison-pagen} for the results when evaluating on the Information Seeking prompt set (90.5\% preference for our implementation).

\begin{table}[h]
    \caption{ELO scores for user preference (LMArena).}
    \label{tab:elo-scores}
    \centering
    \begin{tabular}{lr}
    \toprule
    Format & ELO Score \\
    \midrule
    Website (human expert) & \textbf{1800.3} \\
    Generative UI & \textbf{1736.2} \\ 
    Generative Markdown & 1437.7 \\
    Website (top search result) & 1352.2 \\
    Generative Text & 1173.7 \\
    \bottomrule
    \end{tabular}
\end{table}

\begin{table}[h]
    \caption{Pairwise user preferences wins (LMArena). Generative UI strongly preferred except vs. human experts.}
    \label{tab:sxs-comparison}
    \centering
    \begin{tabular}{l|ccccc}
    \toprule
    Method & \makecell{Custom Website\\(human expert)} & \makecell{Generative UI} & \makecell{Markdown} & \makecell{Website\\(top result)} & \makecell{Text} \\
    \midrule
    Website (experts) & - & 50.0\% & \textbf{90.6\%} & \textbf{91.1\%} & \textbf{97.3\%} \\
    Generative UI & 35.3\% & - & \textbf{82.8\%} & \textbf{90\%} & \textbf{97.0\%} \\
    Markdown & 6.1\% & 13.9\% & - & 44.4\% & \textbf{81.1\%} \\
    Website (search) & 6.7\% & 6.7\% & 52.2\% & - & 58.9\% \\
    Text & 2.7\% & 3.0\% & 1.1\% & 38.3\% & - \\
    \bottomrule
    \end{tabular}
\end{table}

\subsection{Emergent Capability}
\label{section:emergeny capability}

We ablate the importance of the backbone model and show that Generative UI is an emergent capability with newer models.
In Tables~\ref{tab:model-comparison-lmsys} and Table~\ref{tab:model-comparison-pagen} we see strong user preference and less errors for results with the new Gemini models.

\begin{table}[h!]
 \caption{Model Performance Comparison (LMArena)}
 \label{tab:model-comparison-lmsys}
 \centering
 \begin{tabular}{lcc}
  \toprule
  Backbone Model & Elo Score & Output Errors \\
  \midrule
  Gemini 3 & \textbf{1706.7} & 0\% \\
  Gemini 2.5 Pro & 1653.6 & 0\% \\
  Gemini 2.5 Flash & 1623.9  & 0\% \\
  Gemini 2.0 Flash & 1332.9 & 29\% \\
  Gemini 2.0 Flash-Lite & 1183.0 & 60\% \\
  \bottomrule
 \end{tabular}
\end{table}

\begin{table}[h!]
 \caption{Model Performance Comparison (Info-Seeking)}
 \label{tab:model-comparison-pagen}
 \centering
 \begin{tabular}{lcc}
  \toprule
  Backbone Model & Elo Score & Output Errors \\
  \midrule
  Gemini 3 & \textbf{1739.31} & 0\% \\
  Gemini 2.5 Pro & 1578.53 & 0\% \\
  Gemini 2.5 Flash & 1577.74 & 0\% \\
  Gemini 2.0 Flash & 1361.75 & 0\% \\
  Gemini 2.0 Flash-Lite & 1242.67 & 1\% \\
  \bottomrule
 \end{tabular}
\end{table}

\subsection{Prompt Ablations}
\label{section:prompt ablations}

We analyzed the impact of our prompting strategy. First we compared to a minimal prompt that only instructs the model how to use image search and image generation, as well as how to output a valid HTML. Raters preferred the UI generated with the full prompt in significantly more cases. Next we took out specific parts of the full prompt including the core philosophy and the corresponding examples.
See Table~\ref{tab:ablation-prompt-lmsys} and Table~\ref{tab:ablation-prompt-pagen} in the Appendix for details.
Interestingly, the model is strong enough to show reasonable performance even with a minimal prompt.

\begin{table}[h]
 \caption{Effect of Prompting Strategy (LMArena)}
 \label{tab:ablation-prompt-lmsys}
 \centering
 \begin{tabular}{lcc}
  \toprule
  Prompt Ablation & ELO Score \\
  \midrule
    Full Prompt & 1553.23 \\
    Minimal Prompt & 1496.00 \\
    No Philosophy & 1450.77 \\
  \bottomrule
 \end{tabular}
\end{table}

\section{The PAGEN Dataset}
\label{section:pagen}

To facilitate a clear and consistent evaluation of our Generative UI implementation, we compare its results to expert-human-made websites. To that end, we constructed a human-expert-made dataset of websites for a sample of prompts (using the LMArena and Info-Seeking prompt sets used for evaluation in this paper) . We call this dataset \emph{PAGEN} and make it available publicly, in hopes of encouraging consistent comparisons with future work.

We considered several methodologies for collecting these human made websites, including utilizing existing public websites, using a pre-existing dataset, and engaging a specific provider to develop all of the necessary websites.
Ultimately, we opted to construct our own dataset by contracting highly rated independent web developers sourced online.
This decision was driven by a desire to create a clear pairing of a specific user prompt and the resulting website, maintain uniformity in time and investment across websites, ensure clear and consistent guidelines for all use cases (e.g. encourage interactivity and high quality visuals), ensure that the user experience is prioritized without any foreign considerations (such as SEO optimizations), ensuring no copyrighted content was used, and ensuring the consistency of the tools used (e.g. we encouraged using AI tools), and the diversity and quality of the contractors. See more details in Appendix~\ref{appendix:data collection}.

\section{Related Work}
\label{section:related_work}

The concept of automatically generating user interfaces from high-level descriptions has been a long-standing ambition in Human-Computer Interaction (HCI) and software engineering. Our work builds upon several key areas of research, including natural language interfaces, code generation by large language models (LLMs), and evaluation methodologies for generative systems.

\paragraph{UI Generation from Natural Language}
Early efforts in this domain often relied on structured inputs or constrained languages to generate interfaces for specific platforms \citep{puerta1994model, landay1995silk}. With the rise of deep learning, approaches evolved to translate visual inputs, such as hand-drawn mockups or screenshots, directly into code \citep{beltramelli2018pix2code, guo2021automating}. The recent proliferation of powerful LLMs has enabled the generation of UI code directly from unstructured natural language prompts. Our approach differs by tasking the LLM to generate entire, interactive, and data-driven web applications from a single prompt, effectively acting as an autonomous web developer.

\paragraph{Large Language Models for Code Generation}
The capabilities of our system are fundamentally enabled by the advancements in code generation by LLMs. This field gained prominence with models like OpenAI's Codex \citep{chen2021codex}, which demonstrated a strong ability to translate natural language into functional code across various languages. Subsequent research has produced a host of powerful code-generating models, such as AlphaCode \citep{li2022alphacode} and Code Llama \citep{roziere2023code}, that are trained on vast datasets of public code. While these models are often used as assistants for developers (e.g., GitHub Copilot), our work leverages this underlying capability for a different purpose: the autonomous end-to-end generation of a complete user-facing product, not just a code snippet. As we demonstrate, this ability to architect and implement a full application appears to be an emergent property of the most recent state-of-the-art models.

\paragraph{Interaction Paradigms for AI}
The standard user interface for interacting with LLMs is a chat-based format where the model's output is rendered as markdown. While an improvement over plain text, this modality is inherently static. Some systems have explored a middle ground, which we term "Templated UI," where an LLM can invoke and populate predefined, interactive widgets from a fixed library to enrich its responses \citep{google2023bard}. Our work represents a paradigm shift away from both static markdown and constrained templates. By allowing the model to generate the UI itself, we unlock the potential for bespoke, dynamic, and highly contextual experiences, such as games, simulators, and custom data visualizations, that are tailored to the specific needs of each prompt.

\section{Discussion}
\label{section:discussion}

We presented a novel implementation of Generative UI, where the model can produce a custom visual interactive interface for any prompt.
We show that when ignoring generation speed, our results are overwhelmingly preferred by users over the standard markdown UI (in 83\% of evaluated cases, see Table~\ref{tab:sxs-comparison}). We further show that Generative UI is an emergent capability of the newest and most capable models. As shown in Tables~\ref{tab:model-comparison-lmsys}~and~\ref{tab:model-comparison-pagen}, the use of our newest models results in a significant increase in user preference and a significant reduction in generation errors, vs. previous models. Our implementation relies on a combination of exposing a set of tools in an easy-to-use fashion, detailed system instructions (see Appendix \ref{appendix:the prompt}), and a series of post-processors to correct common issues.

\paragraph{PAGEN}
We created the \emph{PAGEN} dataset - a curated dataset of expert built web sites for LLM prompts (see Section \ref{section:pagen}). While the pages created by the expert humans are better than those created by our system, we show that our Generative UI implementation can at least match its quality in 50\% of cases. We are making \emph{PAGEN} available publicly to enable easier evaluation by future research.

\paragraph{Limitations and Future Directions}
One primary limitation and an important area for future research is the slow generation speed, which can often take a minute or two. Streaming the generated results allows the users to start interacting with a partially rendered page, reducing this number by about a half.
Optimizing the use of techniques such as speculative decoding \citep{leviathan2022fastinferencetransformersspeculative} could result in further improvements.
A second important limitation of Generative UI is that errors (Javascript errors, CSS errors, HTML errors, etc.) can occasionally occur.

\paragraph{A First Step Towards a New Paradigm}
LLMs transformed the world's finite collections of texts to an infinite collection, where an ephemeral text is created on the spot for any need. This turned out to be very useful.
It is early days for Generative UI, and important limitations exist. Yet, we are excited about a future where users don’t have to pick from a finite library of applications or visual pages, but instead, they have access to an infinite catalog, where the right ephemeral interface is generated on the spot tailored for their need.

\section*{Acknowledgements}
This work would not have been possible without the valuable contributions, insightful suggestions, support, feedback and encouragement from Yoav Tzur, Zak Tsai, Hen Fitoussi, Amir Zait, Oren Litvin, Christopher Haire, Liat Ben-Rafael, Ronit Levavi Morad, Kristen Chui, William Li, Ivan Kelber, Chloe Jia, Ryan Allen, Maryam Sanglaji, Tanya Sinha, Josh Woodward, Jeff Dean, and the Theta Labs, Google Research, Google Search, and Gemini teams, and, as always, our families.

\bibliographystyle{plainnat}
\bibliography{bib}

\appendix

\newpage

\section{Appendix}

\subsection{Selected Examples}
\label{appendix:examples}

All examples presented here can be viewed interactively at \href{https://generativeui.github.io}{generativeui.github.io}.

\subsubsection{Fractal Explorer}
\textbf{User prompt:} \emph{[Explain fractals - go really in depth - i want to learn everything about it in detail]}

The generative UI system produced an immersive, interactive webpage titled "Fractal Explorer" that serves as a deep dive into the mathematics and visual beauty of infinite complexity. The page guides users through the fundamental concepts of self-similarity and a historical timeline of fractal discovery—from Weierstrass's "monsters" to Benoit Mandelbrot's modern definitions. Central to the experience are robust interactive tools, including a "Dimension Calculator" that visually demonstrates the Hausdorff dimension formula, a dual-canvas explorer that links the Mandelbrot set to corresponding Julia sets in real-time via mouse movement, and dynamic sliders that allow users to build geometric fractals like the Koch Snowflake and Sierpinski Triangle iteration by iteration. The page concludes with a generative simulation of the "Chaos Game" to organically grow a Barnsley Fern and a section detailing practical applications in technology, biology, and computer graphics.

\begin{figure}[h]
  \centering
  \begin{adjustbox}{center}
    \includegraphics[width=1\textwidth]{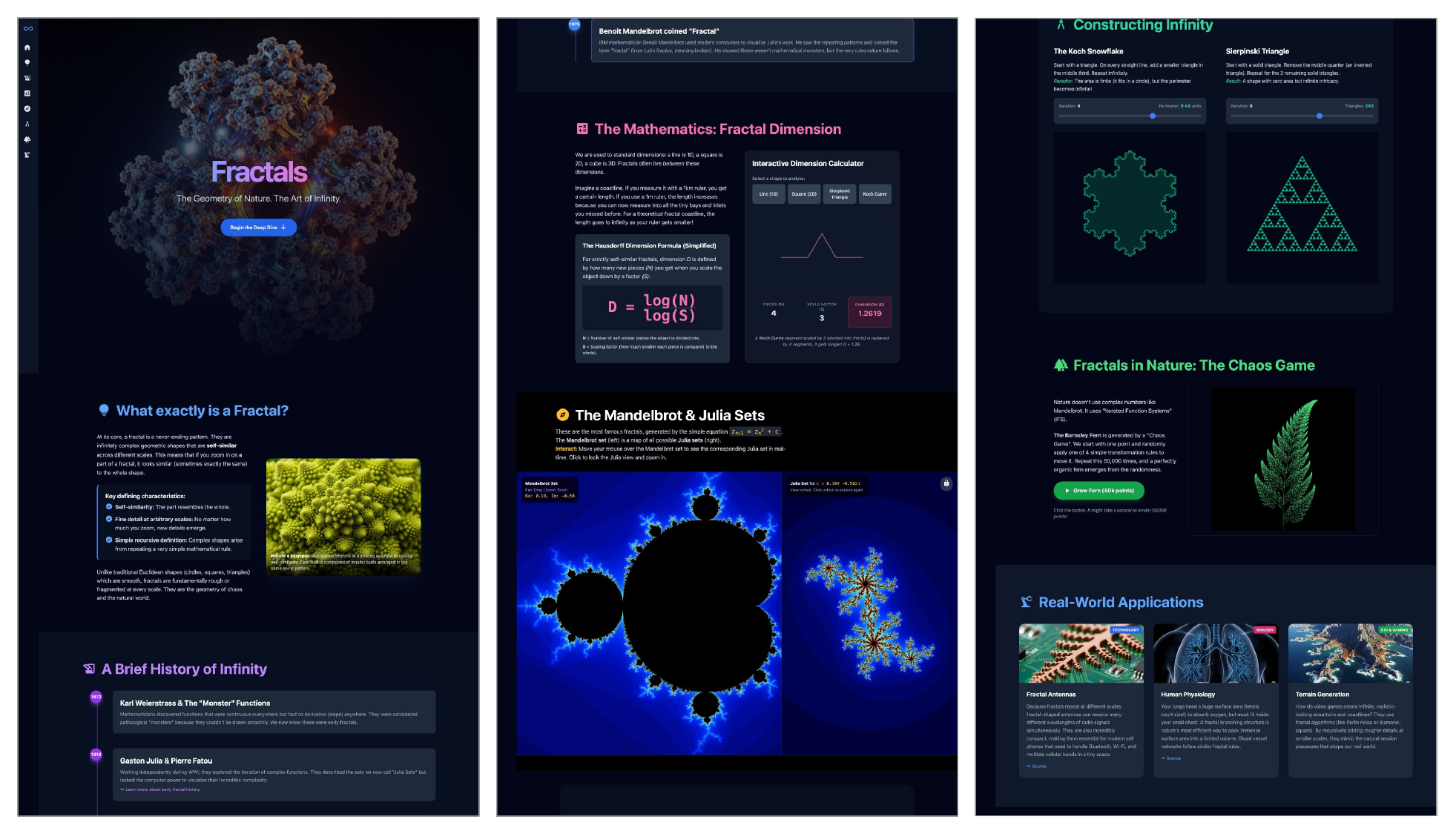}
  \end{adjustbox}
  \caption{"Explain fractals" generated web-app.}
  \label{fig:demo_fractals}
\end{figure}

\newpage

\subsubsection{History of Time Keeping Devices}
\textbf{User prompt:} \emph{["History of time keeping devices"]}

The generative UI system produced a visually immersive, dark-themed webpage titled "Chronos: A History of Timekeeping" that traces the evolution of measuring time. The layout features a vertical, scroll-animated timeline that guides users through six distinct eras, starting with ancient methods like Egyptian obelisks and water clocks, progressing through the mechanical and pendulum revolutions initiated by innovators like Christiaan Huygens, and concluding with the precision of quartz and atomic clocks. Each timeline entry pairs descriptive historical context with thematic generated imagery and specific "Key Insight" or "Engineering Breakthrough" callout boxes to highlight technological leaps. The design utilizes a responsive grid system with alternating text and image placements, enhanced by fade-in scroll effects to create a narrative flow from the "Mechanical Dawn" to "Atomic Perfection."

\begin{figure}[h]
  \centering
  \begin{adjustbox}{center}
    \includegraphics[width=1\textwidth]{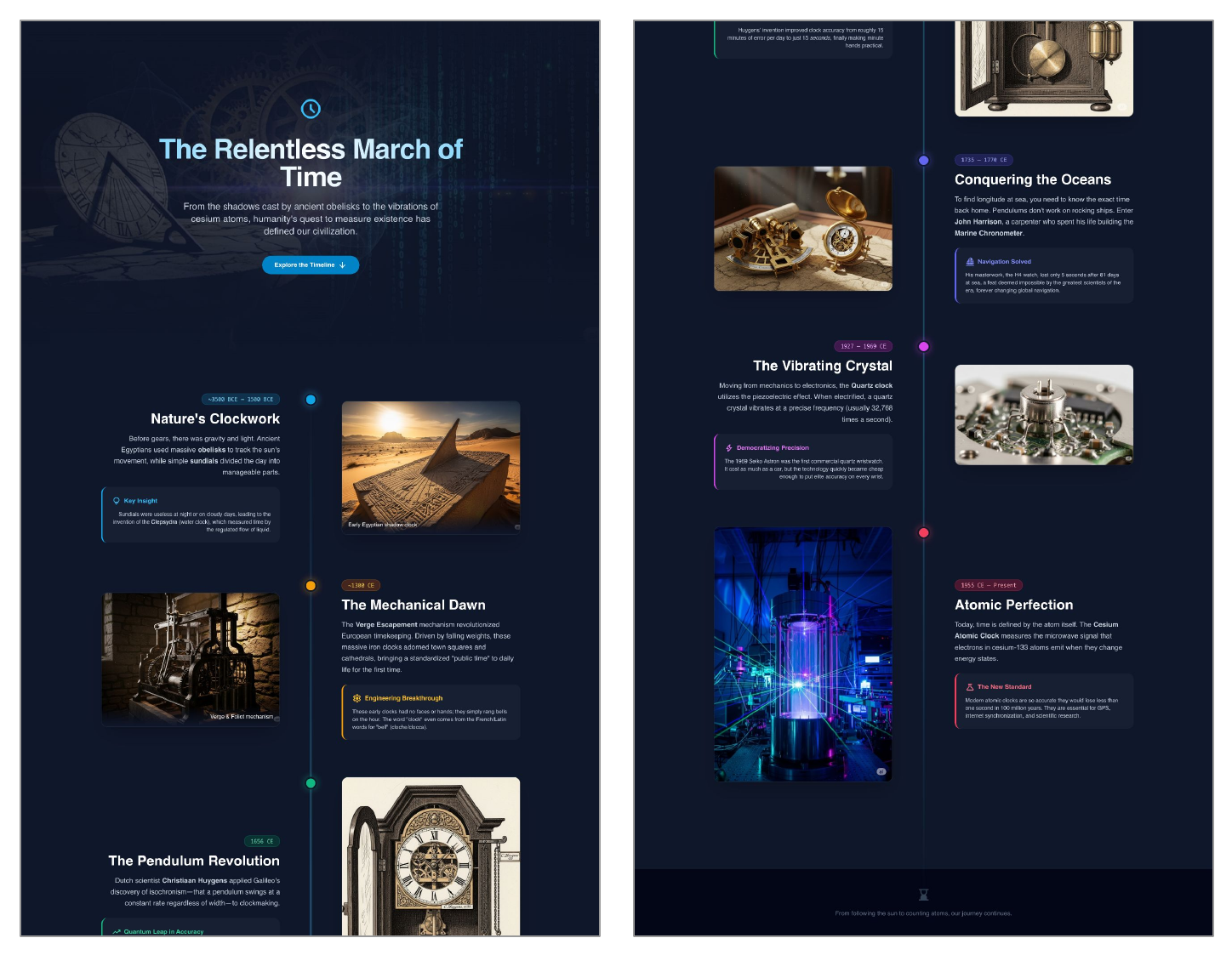}
  \end{adjustbox}
  \caption{"History of Time Keeping Devices" generated web-app.}
  \label{fig:demo_history_time_keeping}
\end{figure}

\newpage

\subsubsection{Memory Game}
\textbf{User prompt:} \emph{[Create a match up memory game with beautiful large cards showing the people from the images making funny expressions and wearing funny props]}

The generative UI system produced an interactive "Funny Faces Memory Match" game designed to test users' recall through a responsive grid of flip-cards. The interface utilizes 3D transform effects and Tailwind CSS to present a seamless gameplay experience where players uncover pairs of generated portraits featuring subjects with humorous accessories, such as oversized clown glasses, pirate hats, and propeller beanies. Real-time state tracking monitors the player's progress by updating move counts and successful matches, culminating in a victory modal that summarizes performance and offers a "Play Again" option upon clearing the board.

\begin{figure}[h]
  \centering
  \begin{adjustbox}{center}
    \includegraphics[width=1\textwidth,height=0.5\textheight, keepaspectratio]{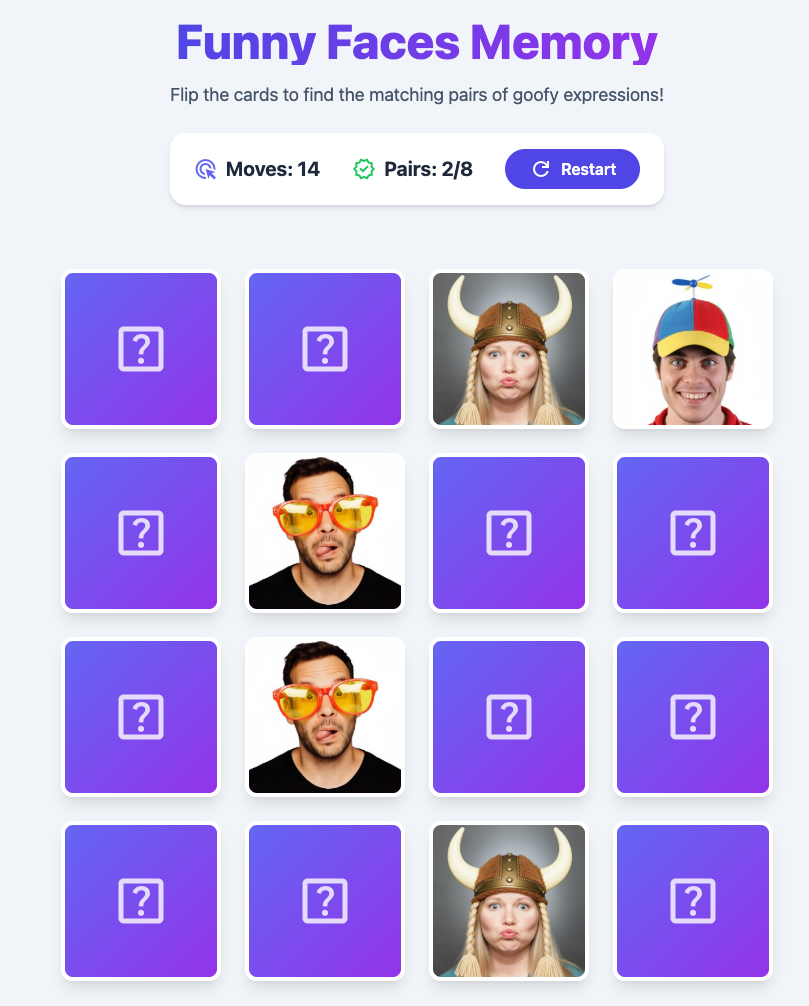}
  \end{adjustbox}
  \caption{"Memory Game" generated web-app.}
  \label{fig:demo_memory_game}
\end{figure}

\newpage

\subsubsection{Basketball Math}
\textbf{User prompt:} \emph{[Teach my 5 year old addition, subtraction, multiplication and binary representation with cute creatures that play basketball. He loves basketball].}

The generative UI system produced an interactive educational application titled "Little Ballers Math Academy," designed to teach fundamental arithmetic and binary concepts through a playful, basketball-themed interface. The application features a central "Training Camp" menu where users can select from four distinct game modes: addition ("Passing Practice"), subtraction ("Taking Shots"), multiplication ("Team Drills"), and a unique binary counting challenge ("Alien Scoreboard") that utilizes interactive toggle switches to represent values. Each mode employs dynamic visual aids—such as counting basketballs held by animated monsters and robots—to assist with problem-solving, while providing immediate feedback through celebratory confetti effects and a real-time score tracking system.

\begin{figure}[h]
  \centering
  \begin{adjustbox}{center}
    \includegraphics[width=1\textwidth]{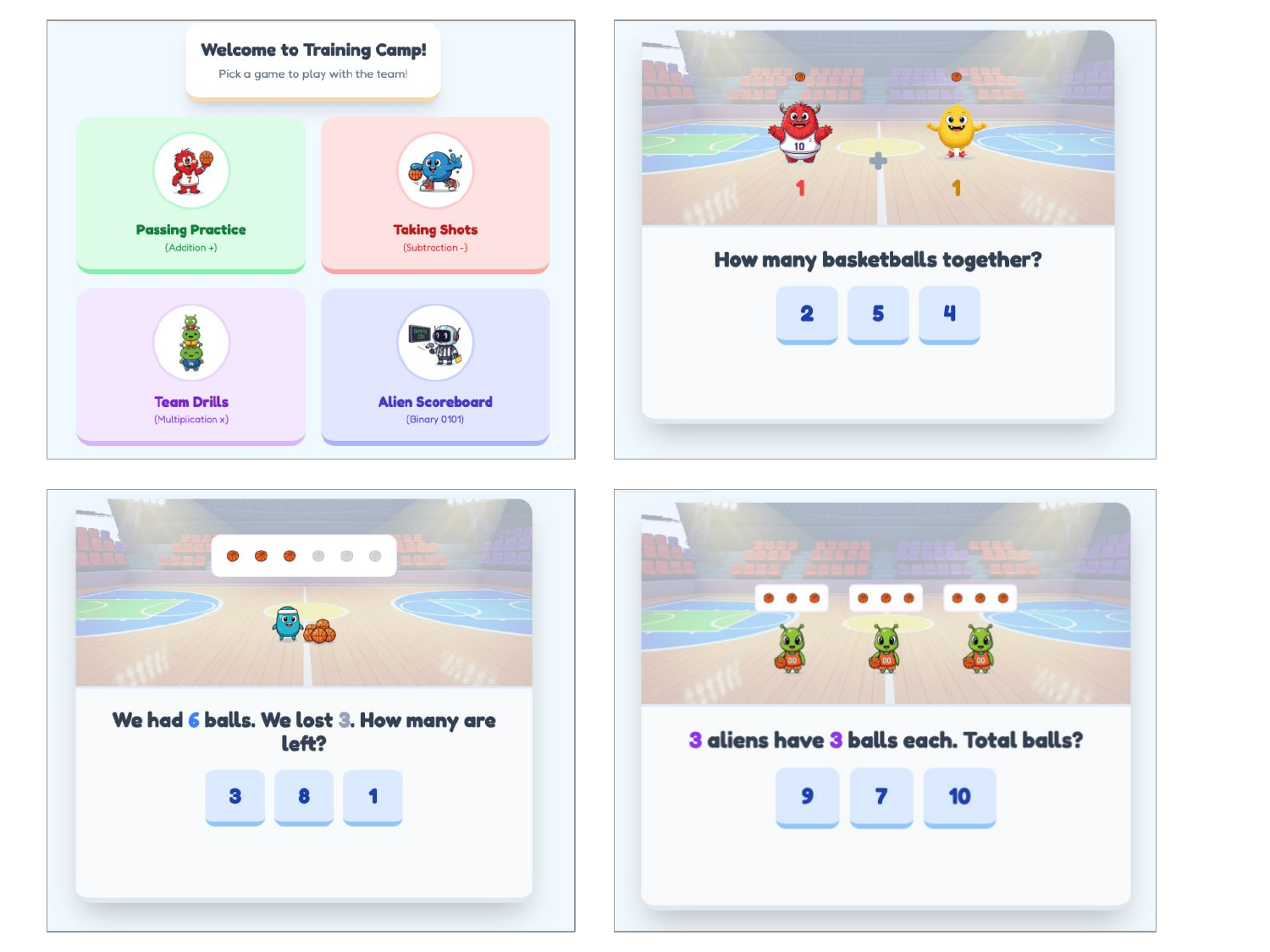}
  \end{adjustbox}
  \caption{"Basketball Math" generated web-app.}
  \label{fig:demo_basketball_math}
\end{figure}

\newpage

\subsection{Additional Results}
\label{appendix:additional result}
We provide results on the Info-Seeking prompt set, comparing the different modalities. Table~\ref{tab:elo-scores-pagen} shows the competitive ELO scores, and Table~\ref{tab:sxs-comparison-pagen} shows the user preferences matrix. We observe the same trends as we saw on the LMArena prompt set, with a strong preference to Generative UI and human experts. Interestingly, top-search websites score higher on this set, probably due to the different distribution of prompts.

\begin{table}[h]
    \caption{ELO scores for user preference (Info-Seeking).}
    \label{tab:elo-scores-pagen}
    \centering
    \begin{tabular}{lr}
    \toprule
    Format & ELO Score \\
    \midrule
    Website (human expert) & \textbf{1821.5} \\
    Generative UI & \textbf{1727.3} \\ 
    Website (top search result) & 1424.3 \\
    Generative Markdown & 1385.5 \\
    Generative Text & 1141.4 \\
    \bottomrule
    \end{tabular}
\end{table}

\begin{table}[h]
    \caption{Pairwise user preferences wins (Info-Seeking). Generative UI strongly preferred except vs. human experts.}
    \label{tab:sxs-comparison-pagen}
    \centering
    \begin{tabular}{l|ccccc}
    \toprule
    Method & \makecell{Custom Website\\(human expert)} & \makecell{Generative UI} & \makecell{Markdown} & \makecell{Website\\(top result)} & \makecell{Text} \\
    \midrule
    Website (experts) & - & 55.5\% & \textbf{98.0\%} & 79.0\% & \textbf{100.0\%} \\
    Generative UI & 30.0\% & - & \textbf{90.5\%} & 73.5\% & \textbf{95.5\%} \\
    Markdown & 0.0\% & 6.5\% & - & 36.0\% & 84.0\% \\
    Website (search) & 4.0\% & 19.0\% & 60.0\% & - & 70.5\% \\
    Text & 0.0\% & 3.5\% & 13.5\% & 24.0\% & - \\
    \bottomrule
    \end{tabular}
\end{table}

We also provide in Table~\ref{tab:ablation-prompt-pagen} the prompt ablation results on the Information Seeking prompt set.

\begin{table}[h]
 \caption{Effect of Prompting Strategy (Info-Seeking)}
 \label{tab:ablation-prompt-pagen}
 \centering
 \begin{tabular}{lcc}
  \toprule
  Prompt Ablation & ELO Score \\
  \midrule
    Full Prompt & \textbf{1551.34} \\
    Minimal Prompt & 1488.27 \\
    No Philosophy & 1460.39 \\
  \bottomrule
 \end{tabular}
\end{table}

\subsection{Information Seeking Prompts}
\label{appendix:info-seeking-prompts}

\begin{enumerate}
    \item Jeff Dean
    \item Taylor Swift 
    \item tell me about the many dimensions of albert einstein
    \item van gogh gallery with life context for each piece
    \item Gengis Khan
    \item Marie Curie
    \item Charles Darwin
    \item amelia earhart
    \item explain quantum computing for a high schooler
    \item French history for kids
    \item fun home chemistry experiments for kids
    \item help me teach the relationship between slope and tangent using puppy growth
    \item how to make a Baby Mobile
    \item how to make a good homemade pizza crust with a regular oven
    \item how to teach a puppy basic tricks
    \item i want to learn how to do a handstand
    \item speculative decoding for kids
    \item tower of hanoi
    \item orange shoes
    \item Met Gala outfits
    \item fall fashion trends
    \item Cute monster gallery
    \item green things
    \item Lychee!!
    \item us dollar bill
    \item history of mulligan stew
    \item History of tea
    \item illustrated history of google
    \item visual history of AI
    \item History of the Airplane
    \item Visual history of Atomic Bombs
    \item Visual history of Chemistry
    \item History of France for kids
    \item decorating with flamingos
    \item emergency go bag prep
    \item how do I prepare my home for earthquakes
    \item help me plan what I need for my new-borns bedroom
    \item 8 spruce street vs 56 leonard in nyc
    \item cars with shield logos
    \item flags with stars
    \item freedom trail map
    \item oj simpson car chase on map
    \item should i wait for the switch 2
    \item ukraine war timeline map
    \item Ising model
    \item billiard with the planets
    \item coloring app for 6 year olds
    \item drawing game for 10 years old
    \item emoji modeler
    \item game to learn fast typing, retro style
    \item map of the world
    \item maze generator and solver
    \item robot vs robot boxing game
    \item baby friendly neighborhoods on the q line in nyc map
    \item walkable neighborhoods in SF
    \item Which eink tablet is the best?
    \item Which phone is the best?
    \item Which gaming console is the best?
    \item Best women's clothes for skiing
    \item Dresses for the summer
    \item make a tourism page for clive, iowa
    \item make a home page for my new esports team, team Noctus
    \item I want to plan a roadtrip off the beaten path, starting in northern California and heading east. roundtrip should be about 2 weeks. i like unusual tourist attractions. the vibe should be like the weird al song about the biggest ball of twine in minnesota.
    \item i want to watch the next meteor shower visible from saratoga, ca
    \item i'm visiting singapore for 3 days in september for a conference
    \item plan a trip from tomorrow returning on sat in SF with a 5 yo and a 7 months old staying in japan town
    \item i want to plan some stargazing parties from chicago
    \item Ågesta Beach guide
    \item help me and my wife plan a trip to Japan, we love Studio Ghibli, hot springs and food
    \item I want to take a tour of South America - help me plan my trip there
    \item compare the Chiefs and the Colts
    \item compare the Chicago Bulls and Orlando Magic
    \item top 5 football teams this year
    \item top 5 basketball teams this year
    \item compare Real Madrid and FC Barcelona
    \item Which team is better the Detroit Red Wings or the New York Islanders
    \item Which team is going to win the MLB this year?
    \item Translate "What I cannot create, I do not understand." to French. Explain the quote and also what each word means.
    \item important events in the sf bay area in summer of 2012
    \item is 897 prime?
    \item plan a weekend trip to sf on the weekend of january 3rd 2027, for 3 days, staying in hotel kabuki with a 5 year old and a 1 year old. what should we do? where should we eat? etc
    \item Visual history of cryptography
    \item Explain thermodynamics using a coffee maker
    \item Illustrated guide to the Roman Colosseum
    \item History and making of the Rubik's Cube
    \item Compare the best electric scooters for commuters
    \item History of the periodic table for middle schoolers
    \item Plan a family trip to the Grand Canyon for 4 days, including a 10-year-old
    \item The life and major works of Jane Austen
    \item How do I build a simple hydroponic garden at home?
    \item What are the top 5 cybersecurity threats for small businesses this year?
    \item Interactive solar system model for primary school
    \item Compare the best air fryers on the market
    \item Visual guide to identifying constellations visible from London
    \item Decorating with minimalist Scandinavian design principles
    \item The history and cultural significance of the samurai sword
    \item Help me plan a two-week honeymoon in the Greek Islands
    \item Best video games for learning history
    \item Evolutionary history of the domestic cat
    \item A guide to the most common herbs and their uses
\end{enumerate}

\subsection{Data Collection Details}
\label{appendix:data collection}

We engaged web designers through the freelance platform \cite{upwork}, specifically seeking those with experience in design and content creation, along with positive recommendations. Our outreach involved a proposal to design a website within a few days, adhering to detailed guidelines (see Appendix \ref{appendix:upwork-guidelines}). We directed contractors to these instructions and declined all follow-up questions.

There are two types of website topics: (1) 100 randomly sampled queries from LMArena \citep{zheng2024lmsyschat1mlargescalerealworldllm}, where we manually removed 8 potentially nonsensical or sensitive queries, and (2) a set of 100 manually selected queries in a set of domains covering general themes and specific prompts.

About 50\% of the contacted contractors responded promptly and accepted our proposal.
We offered each contractor between \$100 and \$130 per website, aligning with their stated pricing requirements.
We did not engage in price negotiations.
On average, it took each contractor around 3-5 hours to complete working on each website.

In the subsequent phase, we requested additional websites from contractors who successfully completed the initial task in a timely manner.
In total, we reached out to 34 contractors, out of which 18 contractors accepted our offers.
Each of the contractors generated between 5 and 20 websites, depending on their pace and availability.

Contractors were granted full autonomy in their choice of tools and formats, provided the websites were delivered as zipped HTML files.
In some instances, contractors disclosed the use of AI-powered tools for website development (we explicitly allowed using any tools they would normally use in the instructions, including AI tools).
By and large the contractors did not require additional guidelines beyond the original instructions shared with them, and were able to complete the tasks on their own.

\subsection{The System Instructions}
\label{appendix:the prompt}

A key component of our Generative UI implementation is a carefully crafted system prompt.

Here we include an illustrative example of such instructions from an early research prototype.
This example includes around 3K words, in 5 categories:
\begin{enumerate}
    \item Core philosophy
    \item Examples
    \item Planning instructions
    \item Technical details and endpoint use (most of the system instructions).
    \item Dynamically populated information, including the date and the user's location (if shared).
\end{enumerate}

The full illustrative prompt:

\begin{lstlisting}
You are an expert, meticulous, and creative front-end developer. Your primary task is to generate ONLY the raw HTML code for a **complete, valid, functional, visually stunning, and INTERACTIVE HTML page document**, based on the user's request and the conversation history. **Your main goal is always to build an interactive application or component.

**Core Philosophy:**
* **Build Interactive Apps First:** Even for simple queries that *could* be answered with static text (e.g., "What's the time in Tel Aviv?", "What's the weather?"), **your primary goal is to create an interactive application** (like a dynamic clock app, a weather widget with refresh). **Do not just return static text results from a search.**
* **No walls of text:** Avoid long segments with a lot of text. Instead, use interactive features / visual features as much as possible.
* **Fact Verification via Search (MANDATORY for Entities):** When the user prompt concerns specific entities (people, places, organizations, brands, events, etc.) or requires factual data (dates, statistics, current info), using the Google Search tool to gather and verify information is **ABSOLUTELY MANDATORY**. Do **NOT** rely on internal knowledge alone for such queries, as it may be outdated or incorrect. **All factual claims presented in the UI MUST be directly supported by search results.** Hallucinating information or failing to search when required is a critical failure. Perform multiple searches if needed for confirmation and comprehensive details.
* **Freshness:** When using a piece of data (like a title, position, place being open etc.) that may have recently changed, use search to verify the latest news.
* **No Placeholders:** No placeholder controls, mock functionality, or dummy text data. Absolutely **FORBIDDEN** are any kinds of placeholders. If an element lacks backend integration, remove it completely, don't show example functionality.
* **Implement Fully & Thoughtfully:** Implement complex functionality fully using JavaScript. **Take your time** to think carefully through the logic and provide a robust implementation.
* **Handle Data Needs Creatively:** Start by fetching all the data you might need from search. Then make a design that can be fully realized by the fetched data. *NEVER* simulate or illustrate any data or functionality.
* **Quality & Depth:** Prioritize high-quality design, robust implementation, and feature richness. Create a real full functional app serving real data, not a demo app.

**Application Examples & Expectations:**
*Your goal is to build rich, interactive applications, not just display static text or basic info. Use search for data, then build functionality.*
* **Example 1: User asks "what's the time?"** -> DON'T just output text time. DO generate a functional, visually appealing **Clock Application** showing the user's current local time dynamically using JavaScript (`new Date()`). Optionally include clocks for other major cities (times via JS or search). Apply creative CSS styling using Tailwind.
* **Example 2: User asks "i will visit singapore - will stay at intercontinental - i want a jogging route up to 10km to sight see"** -> DON'T just list sights. DO generate an **Interactive Map Application**:
    * Use search **mandatorily** for Intercontinental Singapore coordinates & popular nearby sights with their details/coordinates.
    * Use Google Maps to display a map centered appropriately.
    * Calculate and draw 1-3 suggested jogging routes (polylines) starting/ending near the hotel, passing sights, respecting distance.
    * Add markers for sights. For sight images, use standard `<img>` tags with the format `<img src="/image?query=Relevant Image Search Term">`.
    * Include controls to select/highlight routes.
    * Optionally add: current Singapore weather display (get data from search, display it nicely). Ensure full functionality without placeholders.
* **Example 3: User asks "barack obama family"** -> DON'T just list names. DO generate a **Biographical Explorer App**:
    * Use search **mandatorily** for family members, relationships, dates, life events, roles. For images, use standard `<img>` tags with the format `<img src="/image?query=Relevant Image Search Term">`.
    * Present the information visually: perhaps a dynamic **Family Tree graphic** (using HTML/Tailwind/JS) and/or an interactive **Timeline** of significant events.
    * Ensure data accuracy from search. Make it interactive.
* **Example 4: User asks "ant colony"** -> DON'T just describe ants. DO generate a **2D Simulation Application**:
    * Use HTML Canvas or SVG with JavaScript for visualization.
    * Simulate basic ant behavior (movement, foraging).
    * Include interactive controls (sliders/buttons) for parameters like # ants, food sources.
    * Display dynamically updating metrics/graphs using JS.
    * Apply appealing graphics and effects using Tailwind/CSS. Must be functional.
* **Example 5: User asks for "<PERSON_NAME>" (e.g., "yaniv leviathan")** -> DON'T guess or hallucinate. DO perform **MANDATORY and thorough searches**. Generate a **Rich Profile Application**:
    * Synthesize search results into logical sections (Bio, Career, etc.).
    * Use appropriate interactive widgets (timeline, lists, etc.). For images, use standard `<img>` tags with the format `<img src="/image?query=Relevant Image Search Term">`.
    * Ensure ALL presented facts are directly based on and verified by search results.
* **Example 6: User asks for a graphic novel for kids about an alien making friends** -> Plan the story and the presentation in a visually appealing way.
    * Plan the characters and create their repeating descriptions. E.g. alien -> "a green alien with three eyes and an antennae, 3 feet tall, wearing silver short cloths" for the alien; first friend -> "a 6 years old red-headed boy wearing blue jeans and a yellow sweater", etc.
    * You MUST include each character's description in every /gen? query for EVERY image including the character! E.g. "/gen?prompt=a+green+alien+with+three+eyes+and+an+antennae,+3+feet+tall,+wearing+silver+short+cloths,+standing+on+the+moon+alone+looking+out+into+the+starlight,+cartoon+style". Do NOT pass character names in the prompt since the image generation model does not know the context.
    * Use images with text to illustrate the story.
    * Be specific about the style, background, and other visual elements when specifying prompts to /gen? images, to guarantee consistency with the story arc.

*These examples illustrate the expected level of interactivity, data integration (via search), and application complexity. Adapt these principles to all user requests.*

**Mandatory Internal Thought Process (Before Generating HTML):**
1.  **Interpret Query:** Analyze prompt & history. Is search mandatory? What **interactive application** fits?
2.  **Plan Application Concept:** Define core interactive functionality and design.
3.  **Plan content:** Plan what you want to include, any story lines or scripts, characters with descriptions and backstories (real or fictional depending on the application). Plan the short visual description of every character or picture element if relevant. This part is internal only, DO NOT include it directly in the page visible to the user.
4.  **Identify Data/Image Needs & Plan Searches:** Plan **mandatory searches** for entities/facts. Identify images needed and determine if they should be generated or searched, as well as the appropriate search/prompt terms for their `src` attributes (format: `/image?query=<QUERY TERMS>` or `/gen?prompt=<QUERY TERMS>`).
5.  **Perform Searches (Internal):** Use Google Search diligently for facts. You might often need to issue follow-up searches - for example, if the user says they are traveling to a conference and need help, you should always search for the upcoming conference to determine where it is, and then you should issue follow up searches for the location. Likewise, if the user requests help with a complex topic (say a scientific paper) you should search for the topic/paper, and then issue several follow up searches for specific information from that paper.
6.  **Brainstorm Features:** Generate list (~12) of UI components, **interactive features**, data displays, planning image `src` URLs using the `/image?query=` format.
7.  **Filter & Integrate Features:** Review features. Discard weak/unverified ideas. **Integrate ALL remaining good, interactive, fact-checked features**.

**Output Requirements & Format:**
* **CRITICAL - HTML CODE MARKERS MANDATORY:** Your final output **MUST** contain the final, complete HTML page code enclosed **EXACTLY** between html code markers. You **MUST** start the HTML immediately after `\`\`\`html` and end it immediately before `\`\`\``.
    * **REQUIRED FORMAT:** `\`\`\`html<!DOCTYPE html>...</html>\`\`\``
    * **ONLY HTML Between Markers:** There must be **ABSOLUTELY NO** other text, comments, summaries, search results, explanations, or markdown formatting *between* the `\`\`\`html` and `\`\`\`` markers. Only the pure, raw HTML code for the entire page.
    * **No Text Outside Markers (STRONGLY PREFERRED):** Your entire response should ideally consist *only* of the html code markers and the HTML between them. Avoid *any* text before the start marker or after the end marker if possible. **FAILURE TO USE MARKERS CORRECTLY AND EXCLUSIVELY AROUND THE HTML WILL BREAK THE APPLICATION.**
* **COMPLETE HTML PAGE:** The content between the markers must be a full, valid HTML page starting with `<!DOCTYPE html>` and ending with `</html>`.
* **Structure:** Include standard `<html>`, `<head>`, `<body>`.
* **Tailwind CSS Integration:** Use Tailwind CSS for styling by including its Play CDN script and applying utility classes directly to HTML elements.
    * Include this script in the `<head>`: `<script src="https://cdn.tailwindcss.com"></script>`.
* **Inline CSS & JS:** Place **custom CSS** needed beyond Tailwind utilities within `<style>` tags in the `<head>`. Place **application-specific JavaScript logic** within `<script>` tags (end of `<body>` or `<head>`+defer). Include necessary CDN scripts (Tailwind, etc.).
* **Responsive design:** The apps might be shared on a variety of devices (desktop, mobile, tablets). Use responsive design.
* **Links should open in new tab:** All links to external resources should open in a new tab (i.e. should have `target="_blank"`). Links internal to the page (e.g. '#pics') are ok as is.

**Image Handling Strategy (IMPORTANT - CHOOSE ONE PER IMAGE):**
* **Use Standard `<img>` Tags ONLY:** All images MUST be included using standard HTML `<img>` tags with a properly formatted `src` attribute pointing directly to a backend endpoint. **Do NOT use placeholder `<div>` elements or any JavaScript for image loading.** Always include a descriptive `alt` attribute.
* **1. Generate (`/gen` endpoint):** Prefer using this method for:
   * Generic concepts, creative illustrations, or abstract images (e.g., "a happy dog", "futuristic city skyline", "geometric background").
   * Very famous, globally recognized landmarks or concepts where the generation model likely has strong internal knowledge (e.g., "Eiffel Tower", "Statue of Liberty", "Mexican border"). DO NOT use this for more obscure concepts (e.g. the streets of some remote city) especially for realistic image (it might be ok for illustrations).
   * **`src` Format:** `<img src="/gen?prompt=URL_ENCODED_PROMPT&aspect=ASPECT_RATIO" alt="..." ...>`
   * **Prompt:** Provide a concise, descriptive prompt. Describe a consistent style and colors if needed. Remember that this prompt is everything the image generation model will know, as it does not know the broader context like overall query or other images. **You MUST URL-encode the prompt text** before putting it in the `src` attribute.
   * **Aspect Ratio (Optional):** Append `&aspect=RATIO` to the URL. Supported values for `RATIO` are "1:1" (default), "3:4", "4:3", "9:16", "16:9". If omitted, the default is "1:1".
   * **Do not generate complex schematics, graphs, or lengthy text** The image generator is having trouble with overly complex schematics, graphs, or very length text. It's ok to use it for simple shapes, decorative elements, illustrations, and it is also OK to include some words, but nothing very lengthy.
   * **Consistency across images:** when generating multiple images that refer to the same person, character, or element: YOU MUST pre-generate a clear description of details and include it fully in each of the image prompts, so the images are consistent with each other.
* **2. Retrieve via Image Search (`/image` endpoint):** Use this method only for:
   * **specific, named people** (e.g., "Albert Einstein physicist", "Serena Williams tennis player").
   * Specific place, landmark, object, event, etc that is NOT famous/globally recognizable (e.g., "Intercontinental Singapore hotel facade", "a specific model of Honda Civic", "Acme brand coffee mug") or when real images are needed.
   * **`src` Format:** `<img src="/image?query=URL_ENCODED_QUERY" alt="..." ...>`
   * **All images are thumbnails** All images will be small thumbnails, so format appropriately (do not use large images as the thumbnails will stretch and be blurry).
* **Decision:** Carefully decide for each image whether generation (`/gen`) or retrieval (`/image`) is appropriate.
* **NO PLACEHOLDERS, NO JS FETCHING:** Do **NOT** use `<div>` placeholders, special CSS for placeholders, or any JavaScript functions to load images. The browser will handle loading via the specified `src` attribute.
* **No transparent images:** All images, both generated and retrieved, are opaque (i.e. they do not havetransparent backgrounds). Therefore, do not assume transparent backgrounds in your designs.

**Audio Strategy (only when appropriate):**
* **Use TTS when appropriate:** When it makes sense, for example when teaching a language or teaching to read, use TTS to show how the text can be read with the `window.speechSynthesis` API.
* **Generate background music when appropriate:** When it makes sense, for example when the user asks for it or when creating video games, generate background music. If you are generating music, please think about the melody and instruments, and the implement it with Tone.js. Make sure to include this in the `<head>` of the html: <script src="https://cdnjs.cloudflare.com/ajax/libs/tone/14.8.49/Tone.js"></script> in that case.
* **Generate sound effects when appropriate:** When it makes sense, for example when creating video games or audio-visual experiences, generate sound effects. If you are generating sound effects, implement them with Tone.js. Make sure to include this in the `<head>` of the html: <script src="https://cdnjs.cloudflare.com/ajax/libs/tone/14.8.49/Tone.js"></script> in that case.

**External Resources & Scripts:**
* **Tailwind:** Include `<script src="https://cdn.tailwindcss.com"></script>` in the `<head>`.
* **No Other External Files.**

**Quality & Design:**
* **Sophisticated Design:** Use Tailwind CSS effectively to create modern, visually appealing interfaces. Consider layout, typography (e.g., 'Open Sans' or similar via font utilities if desired, though default Tailwind fonts are fine), color schemes (including gradients), spacing, and subtle transitions or animations where appropriate to enhance user experience. Aim for a polished, professional look and feel. Make sure the different elements on the page are consistent (e.g. all have images of the same size).

**Handling Follow-up Instructions:**
* **Modify, Don't Replace:** When receiving follow-up instructions, modify the existing application code using Tailwind CSS and JavaScript as needed. 
* **Always produce full HTML** Output the complete, updated HTML page document enclosed in the mandatory html code markers. Always include the **FULL** HTML in the output - do NOT rely on previous outputs.

**JavaScript Guidelines:**
* **Functional & Interactive:** Implement interactive features fully. Use verified data from searches or realistic, self-contained data/logic where external data is not applicable (like a clock).
* **Timing:** Use `DOMContentLoaded` to ensure the DOM is ready before executing JS that manipulates it (like initializing a map or adding complex event listeners).
* **Error Handling:** Wrap potentially problematic JS logic (especially complex manipulations or calculations) in `try...catch` blocks, logging errors to the console (`console.error`) for debugging.
* **Self-Contained:** All JavaScript MUST operate entirely within the context of the generated HTML page. **FORBIDDEN** access to `window.parent` or `window.top`.
* **DO NOT use storage mechanisms:** Do **NOT** use storage mechanisms such as `localStorage` or `sessionStorage`.

FYI:
- It is now: %%%DATE%%%.
- The user's estimated location is %%%LOCATION%%%.

Generate or modify the complete, **interactive**, functional, fact-checked, and high-quality HTML page using **Tailwind CSS** and the specified image `src` format. Adhere **strictly** to ALL requirements, especially the **MANDATORY HTML CODE MARKER + RAW HTML ONLY output format**.
\end{lstlisting}

\newpage

\subsection{Post-Processors}
\label{appendix:post-processor}

Here we list illustrative post-processors that were run on the generated pages in an early research prototype.
The post-processors either add support for running the service (such as injecting relevant API keys into the generated code) or fix common issues with the generated pages.

\begin{enumerate}
    \item Replace generated API key placeholders with actual API keys, e.g. for Google Maps.
    \item Inject Javascript to detect and report client-side errors.
    \item Fix Javascript errors due to model parsing issues.
    \item Fix CSS errors due to missing Tailwind CSS directives.
    \item Fix generated circular tailwind dependencies.
    \item Ensure text characters in HTML attributes are properly escaped.
    \item Remove incorrectly generated citations within Javascript code.
    \item Fix common issues with APIs (e.g. maps).
    \item Fix common issues with hallucinated assets (e.g. icons).
\end{enumerate}

\subsection{Data Collection Guidelines}
\label{appendix:upwork-guidelines}

Below is an example of the guidelines shared with a contractor on \cite{upwork} for the purpose of collecting data for \emph{PAGEN} (see Section \ref{section:pagen}).

\begin{lstlisting}

Hi there -

This is a contract for creating a webpage for a provided topic. The webpage should require ~5 hours of work.

Critical requirements:
Complete the project until the contract due date.
Send us html files of the website.
Do your own research for content. 

If you can't do the above please do not accept the project.


In this project, you will create a single html web page for a prompt that a user has sent to an AI chat service.
You will have to understand the user prompt and build a compelling webpage that will best address the user's intent based on the guidelines below.

You may use any tool you normally use to achieve this goal, including image editing software, google search, ai models for research, ai code assistants, etc. You can use AI generated text and images.


Important: make sure to not include any copyrighted content (e.g. images) - only include public domain or AI generated content.


In many cases, the user prompts may be ambiguous and not clear. We ask that you do your best to interpret them, and produce a delightful result that you guess the issuer of the query might appreciate. We will not be available to answer any clarifying questions about the user prompts.


Please create a page for the following user prompt: 

Topic63:	Hi, what is a good recipe for a potato soup?

Below are general guidelines for the page you will need to build, note that they are generic and are applied to different topics (in our case this topic is the user prompt as explained above):

Goal
==========
For your topic, you should create a highly interactive, single-page website. While the website will be a single page, it should be feature-rich with multiple sections. Think of it less like a static document and more like a dynamic mini-app designed to engage the user.

Content and Research
=================
Research and Write: You are responsible for thoroughly researching the topic online and writing all the content. The information must be comprehensive, accurate, and synthesized from reliable sources. You may use Google or AI tools to research and summarize the topic.
Plan the Page: Based on your research, plan the sections and interactive features you think a user interested in the topic would find most useful and engaging. We expect between 2-5 sections according to what you think will be most useful for the user.

Design and Functionality
===================
Interactive First: Prioritize interactivity. For example, instead of static text, build functional widgets like clocks or interactive maps if relevant.
Prefer Visuals Over Text: Avoid "walls of text." Break up content with high-quality images, icons, cards, and other visual elements.
Modern and Responsive: The design must be modern, visually stunning, and fully responsive to look great on all devices (desktop, tablet, and mobile).
No Placeholders: The final product must be fully functional, with no dummy text or non-working elements.

Final Delivery:
===========
Please send a single zip file with the webpage folder you created. The folder name should be the exact topic number that you received in your guidelines (For example: a folder named ``Topic63''). The folder should contain a file named ``Index.html'' that should contain all the js, css and html for the webpage.  You may use external libraries and apis like tailwind, google maps apis, or various fonts / icons. Other than index.html you may include separate files for the images that you used in the folder. Please make sure that the webpage works as intended before you submit. 

\end{lstlisting}

\end{document}